\documentclass[conference]{IEEEtran}

\usepackage[hyphens]{url}
\usepackage{cite}
\usepackage{amsmath,amssymb,amsfonts}
\usepackage{algorithmic}
\usepackage{graphicx}
\usepackage{textcomp}
\usepackage{xcolor}
\def\BibTeX{{\rm B\kern-.05em{\sc i\kern-.025em b}\kern-.08em
    T\kern-.1667em\lower.7ex\hbox{E}\kern-.125emX}}

\AtBeginDocument{%
  \providecommand\BibTeX{{%
    \normalfont B\kern-0.5em{\scshape i\kern-0.25em b}\kern-0.8em\TeX}}}

\usepackage{listings}
\lstdefinelanguage{SQLplus}{
  language     = SQL,
  deletekeywords = {USING}
}

\usepackage{balance}
\begin{document}

\title{Near Data Processing in Taurus Database}

\author{\IEEEauthorblockN{Shu Lin\IEEEauthorrefmark{1}, 
        \space Arunprasad P. Marathe\IEEEauthorrefmark{2},
        \space {Per{-}{\AA}ke Larson}\IEEEauthorrefmark{3},
        \space Chong Chen\IEEEauthorrefmark{4},\\
        \space Calvin Sun\IEEEauthorrefmark{5}, 
        \space Paul Lee\IEEEauthorrefmark{6},
        \space Weidong Yu\IEEEauthorrefmark{7} }
\IEEEauthorblockA{\textit{
Huawei Technologies Canada Co., Ltd., Markham, Ontario, Canada}}
\IEEEauthorblockN{
        \space Jianwei Li\IEEEauthorrefmark{8},
        \space Juncai Meng\IEEEauthorrefmark{9}, 
        \space Roulin Lin\IEEEauthorrefmark{10},
        \space Xiaoyang Chen\IEEEauthorrefmark{11},
        \space Qingping Zhu\IEEEauthorrefmark{12} }
\IEEEauthorblockA{\textit{Huawei Technologies China} }
Email: \{\IEEEauthorrefmark{1}shu.lin,
\IEEEauthorrefmark{2}arun.marathe,
\IEEEauthorrefmark{3}paul.larson,
\IEEEauthorrefmark{4}chongchen,
\IEEEauthorrefmark{5}calvin.sun3,
\IEEEauthorrefmark{6}paul.lee1,
\IEEEauthorrefmark{7}weidong.yu,\\
\IEEEauthorrefmark{8}lijianwei.li,
\IEEEauthorrefmark{9}hw.mengjuncai,
\IEEEauthorrefmark{10}linruolin,
\IEEEauthorrefmark{11}chenxiaoyang12,
\IEEEauthorrefmark{12}zhuqingping1\}@huawei.com
}

\maketitle
\thispagestyle{plain}
\pagestyle{plain}

\begin{abstract}
Huawei's cloud-native database system GaussDB for MySQL
(also known as Taurus) stores
data in a separate storage layer 
consisting of a pool of storage servers.
Each server has considerable compute power 
making it possible to push data reduction operations 
(selection, projection, and aggregation) close to storage.
This paper describes the design and implementation of
near data processing (NDP) in Taurus.
NDP has several benefits: it reduces the amount of data shipped 
over the network; frees up CPU capacity 
in the compute layer; and reduces query run time, thereby
enabling higher system throughput.
Experiments with the TPC-H benchmark (100 GB)
showed that 18 out of 22 queries benefited from NDP;
data shipped was reduced by 63\%; and CPU time by 50\%.
On Q15 the impact was even higher: 
data shipped was reduced by 98\%; CPU time
by 91\%; and run time by 80\%.
\end{abstract}

\begin{IEEEkeywords}
cloud DBMS systems, query processing,
storage virtualization, selection pushdown,
early data reduction, online analytical processing, database
engine architecture
\end{IEEEkeywords}

\section{Introduction}
Applications are increasingly migrating  
to cloud platforms offered by vendors
including Amazon, Microsoft, Google, Alibaba, and Huawei.
Many applications store their data in a relational database,
making relational database services a crucial part
of a cloud platform.

Huawei's cloud platform includes several database offerings
under the unifying brand `{GaussDB}'. {GaussDB} for {MySQL} is a cloud-native 
database service, fully compatible with {MySQL}. 
The underlying technology is called Taurus,
and we will use this term here.

Taurus separates compute and storage. Data is divided into slices 
that are distributed among a number of multi-tenant Page Stores. The {DBMS} frontend, 
where all query processing occurs,  
is a slightly modified version of {MySQL} 8.0. A summary description 
of the architecture is provided in Section \ref{taurus_overview_section}.
A detailed description can be found in \cite{taurus_paper}.

Query processing in {MySQL} is designed for transactional workloads, 
dominated by simple queries and short transactions. 
{MySQL} performs poorly on  queries that sift through large
amounts of data~\cite{mysql_orca_integration_paper}.
In a cloud environment where compute and storage
are decoupled, the network is a shared resource that may become
overloaded, so it is important to minimize network load.
Early data filtering can reduce data volume and
network utilization simultaneously. As a result,
CPU load on the frontend server also reduces, thus
enabling higher system throughput.

Relational query processors try to reduce the amount of data 
to process by executing data reduction operators---selection
(filtering), projection, and aggregation---as early as possible.
Near data processing (NDP) goes a step further, 
and pushes down selection, projection, and aggregation to storage nodes. This reduces {CPU} load on the frontend server,
and spreads it over multiple storage nodes.
It also reduces the amount of data shipped from storage
over the network---sometimes dramatically---which may 
reduce query run time substantially. For example, 
on TPC-H~\cite{tpch_benchmark} Q6, network data volume and run time
were reduced by 99\% and 89\%, respectively. 
The experimental results appear in
Section~\ref{experiments_section} of this paper.
NDP is an idea that can be and has been applied 
at many levels of the memory hierarchy, and in different
software systems---for example, cloud storage services
(Amazon S3 Select); database storage servers (Oracle Exadata); 
and SSD controllers (SmartSSD). More detail about prior work is
provided in Section~\ref{related_work_section} on related work.

This paper is about engineering NDP into an existing code
base in an effective, yet minimally disruptive manner. 
Our NDP design and implementation have the following
noteworthy features.
\begin{itemize}
    \item NDP processing is completely encapsulated within and below the InnoDB storage engine---in fact, almost entirely within index scan cursors.
    The {MySQL} query execution layers above the storage engine are unaware of NDP processing.
    \item {MySQL} query execution depends on index scans 
    returning rows in sorted order, and 
    row versions consistent with the scan's read-view 
    (multi-versioning).
    Our implementation ensures that NDP-enhanced scans
    still satisfy these properties.
    \item An NDP scan reads batches of pages, and parallelizes 
    reads across Page Stores. By contrast, a regular InnoDB scan
    does not perform batch reads.
    \item Selection predicates are converted into
    LLVM ~\cite{llvm} intermediate representation (IR) on the compute node. The IR is compiled into architecture-specific native code on storage nodes.
    \item Page Stores treat NDP processing as a best-effort activity to 
         minimize the impact on other Page Store tenants.
        A Page Store is free to ignore an NDP processing request, and return unprocessed database pages. Any remaining NDP processing is completed by InnoDB on the compute node.
    \item Only a subset of the queries undergo NDP processing, and our design
    ensures that non-NDP queries do not suffer any performance penalties due to
    the new `NDP' code path.
 \end{itemize}

The rest of this paper is organized as follows. A brief overview
of the Taurus architecture appears in Section~\ref{taurus_overview_section}.
A high-level overview of the {NDP} solution appears in
Section~\ref{ndp_overview}. The NDP system design is presented
in Section~\ref{ndp_overall_design}, and includes the NDP-related
changes in the query optimizer, {InnoDB} storage engine, and Page Stores.
Details of how NDP accomplishes column projection, predicate evaluation, and aggregation are presented in 
Section~\ref{ndp_implementation_section}.

Taurus combines NDP with the ability to execute query
plans in parallel, and the resulting synergy enables three
levels of parallelism as described in Section~\ref{pq_and_ndp}.
Experimental results are captured in
Section~\ref{experiments_section}. The related work is
described in Section~\ref{related_work_section}, and conclusions
and some future work are mentioned in Section~\ref{conclusion_section}.

\section{Taurus Overview}
\label{taurus_overview_section}
Taurus is a relational database architecture designed 
by Huawei for multi-tenant cloud environments. 
This section contains a brief overview of the design; a more detailed description can be found in~\cite{taurus_paper}.

Taurus separates compute and storage,
and relies only on append-only storage.
Its architecture and replication algorithms result in
higher availability than the traditional quorum-based replication
without sacrificing performance or increasing
hardware costs. The replication algorithms use separate
persistence mechanisms for database logs and pages,
and ensure strong consistency for logs and eventual
consistency for pages to optimize performance and availability. 

\begin{figure}[h]
  \centering
  \includegraphics[height=1.7in,width=\linewidth]{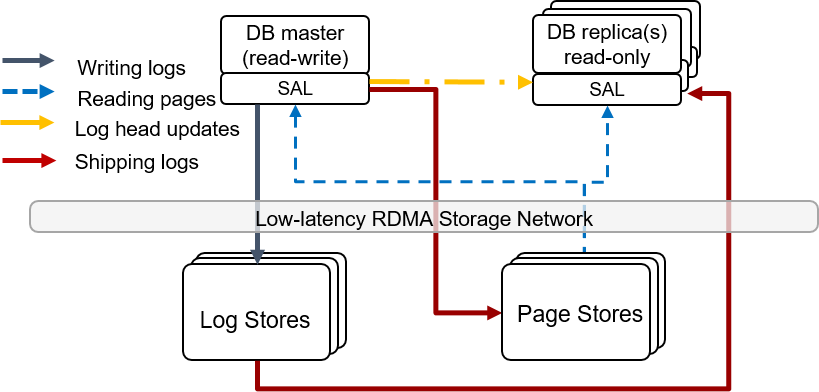}
  \caption{Taurus architecture.}
  \label{fig:taurus_arch}
\end{figure}

As illustrated in Fig.~\ref{fig:taurus_arch}, a Taurus DBMS
consists of four major logical components: database frontends (DB master
and replica nodes); a Storage Abstraction Layer (SAL); Log Stores;
and Page Stores. These components are distributed between two
physical layers: a compute layer and a storage layer, shown
on the two sides of the network layer in Fig.~\ref{fig:taurus_arch}. 
The database is divided into
fixed-size (10 GB) segments called \textit{slices}
that are distributed among multiple Page Stores. Log Stores and Page Stores are multi-tenant services shared by many database servers.

Taurus storage is designed to work with different database frontends:  MySQL, PostgreSQL, and {openGauss}. The frontend layer consists of one master that can serve both read and write queries, and up to 15  read replicas that execute read queries only. 
A frontend server is responsible for accepting incoming connections; optimizing and executing queries; and managing transactions.
All of the updates are handled by the master, whose job is
to make modifications to database pages persistent by
synchronously writing log records, in triplicate,
to durable storage in Log Stores. 
The master also periodically communicates the locations of the
latest log records to all of the read-only replicas
so that they can read the latest log entries,
and update any affected pages in their buffer pools. 

The Storage Abstraction Layer (SAL) is an independent component running on the database server. The SAL isolates the database frontend from the underlying complexity of remote storage; slicing of the database; recovery;
and read replica synchronization. The SAL writes log records to Log Stores;
distributes them to Page Stores; and reads pages from Page Stores.
The SAL is also responsible for creating, managing, and destroying slices in Page Stores; and routing page read requests to Page Stores. 

A Log Store is a service executing in the storage layer responsible for storing log records durably. Once all 
of the log records belonging to a transaction have been made durable, transaction completion can be acknowledged to the client. Log Stores serve two purposes. First and foremost, they ensure the durability of log records. Second, they also serve log records to read replicas so that the replicas can apply the log records to the pages in their buffer pools. 

Page Store servers are also located in the storage layer. A Page Store server hosts slices from multiple database frontends (tenants). 
However, a slice contains  table and
index data from only one database, thereby achieving 
tenant-level data separation.
Each slice is replicated to three Page Stores for durability and availability. The main function of a Page Store is to keep pages up-to-date, and serve read requests from the masters and replicas.
A Page Store receives log records from multiple masters
for the pages it hosts, and applies the log records
to bring pages up-to-date so they are ready to be served.

\section{Life of a query with NDP}
\label{ndp_overview}
This section provides an overview of how an example query undergoes
NDP processing. The NDP design and implementation are described
in detail in sections \ref{ndp_overall_design} and
\ref{ndp_implementation_section}, respectively.
The sample query in Listing~\ref{lst:sample_query}
that computes the average salary of
workers younger than 40 who joined the company
in 2010, is used as an example.

\begin{flushleft}
\begin{minipage}{\linewidth}
\begin{lstlisting}[basicstyle=\footnotesize,frame={tb},language=SQL,breaklines=true,caption={A sample query to demonstrate the effects of NDP.},label={lst:sample_query}]
SELECT AVG(salary)
FROM Worker
WHERE age < 40 AND
  join_date >= DATE '2010-01-01' AND
  join_date < DATE '2010-01-01' + INTERVAL '1' YEAR;
\end{lstlisting}
\end{minipage}
\end{flushleft}

A part of the query's EXPLAIN output describing {MySQL}'s
execution plan with NDP-related information is shown in 
Listing~\ref{lst:ndp_explain_output}.
For brevity, only the relevant information---appearing in the `Extra'
column---is shown.
\begin{flushleft}
\begin{minipage}{\linewidth}
\begin{lstlisting}[basicstyle=\footnotesize,frame=tb,language=SQLplus,breaklines=true,caption={NDP-related information in {MySQL}'s EXPLAIN output.},label={lst:ndp_explain_output}]
Using pushed NDP condition
(((testdb.worker.join_date >= DATE'2010-01-01') AND
  (testdb.worker.join_date < <cache>((DATE'2010-01-01' + INTERVAL '1' YEAR))) AND
  (testdb.worker.age < 40)));
Using pushed NDP columns; Using pushed NDP aggregate
\end{lstlisting}
\end{minipage}
\end{flushleft}

In this example, the entire \texttt{WHERE} clause is pushed into Page Stores,
but this is not always the case.
Residual predicates may remain, 
to be evaluated by the MySQL query executor in the compute node.
Because the query only projects one column out of many in the
\textit{Worker} table, NDP column projection is also chosen. The calculation of \texttt{AVG} is pushed down as well. In short,
the query benefited from NDP fully because all three types
of pushdowns happened, but in general, the three
decisions are taken independently.

NDP processing begins closest to where the data lives:
inside Page Stores.
After applying NDP processing to a page,
the Page Store returns the result as a special NDP-page.
NDP pages may have fewer rows remaining because of predicate filtering; and the rows themselves  
may be narrower (due to NDP column projections) and aggregated (due to NDP aggregation).
NDP pages from one query are unlikely to be of use to other
queries. Accordingly, they are stored in a separate buffer,
and are accessible only to the query that requested the pages.

Selection predicates that have been pushed down to Page
Stores---all of the \texttt{WHERE} conditions for the
query in Listing~\ref{lst:sample_query}---are compiled
into an LLVM bitcode function~\cite{llvm},
and then to native code using just-in-time compilation.
When a Page Store receives a read request for a page,
it first filters the rows by calling the compiled function.
The remaining rows undergo NDP column projection, and only
the columns requested by the query are retained.
Next, partial aggregation is performed, and the sum of salary and the number of rows associated with the sum---using
which \texttt{AVG(salary)} can be computed---are retained. 
The remaining narrowed and aggregated
rows are stored in special NDP pages, and returned to
{MySQL}'s InnoDB storage engine via the SAL.

Next, any residual predicates---none in the example
query---are evaluated by the MySQL query executor,
projection expressions computed, and query results produced.
The query executor orchestrates execution
as before: iterators are initiated top-down in a tree, and
data and result rows percolate bottom-up.

The process described is for a particular query. In the
general case, it is carried out separately for each query block
in a complex query with subqueries. 
In case of an inner query block, the result produced is
consumed by the containing query block.
NDP functionality is largely encapsulated within the
index scan operator, and that operator can appear in any
query block---main or inner---within the query.

The query optimizer in Taurus can produce a parallel query plan, in which multiple workers scan a table concurrently. Each worker scans a portion of the data, and may perform NDP operations in the scan.
\texttt{AVG} is computed by keeping
\texttt{SUM} and \texttt{COUNT} values per thread, 
and a separate ‘leader’ thread then aggregates
the partial values.

\section{Design of the NDP System}
\label{ndp_overall_design}
\begin{figure}[h]
  \centering
  \includegraphics[height=2.5in,width=\linewidth]{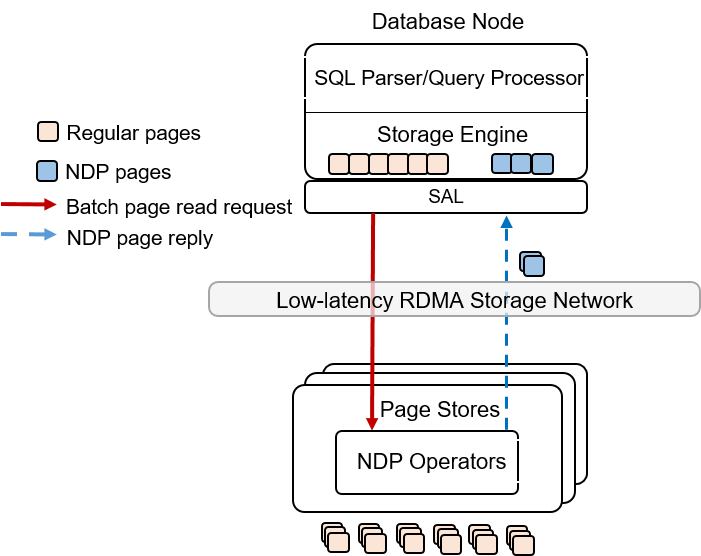}
  \caption{NDP flow and the affected software components}
  \label{fig:taurus_layers}
\end{figure}

This section describes the NDP design in more detail, beginning with
a summary of design goals and constraints.
The changes required to support NDP involve three subsystems: 
the MySQL query optimizer (Section~\ref{ndp_support_in_qo}), the
{InnoDB} storage engine (Section~\ref{ndp_support_in_innodb}),
and Page Stores (Section~\ref{ndp_support_in_page_stores}). Taurus NDP flows and the affected subsystems are shown in  Fig.~\ref{fig:taurus_layers}.

\subsection{Design goals and constraints}
\label{ndp_design_goals_and_constraints}
An important design goal was to minimize the effect
of NDP-related changes to the
software layers above the {InnoDB} storage engine. 
This was achieved by encapsulating NDP processing entirely 
within the index scan operator, and making it invisible to
the operators higher up in a query tree. 

An InnoDB table is always accessed
by scanning an index  (primary or secondary) in forward or reverse order.
Rows are returned in sorted order on the index key, 
and other operators may depend, implicitly or explicitly,
on receiving rows in sorted order. It was important to retain this
property of scans when NDP was enabled. 

A Page Store is a multi-tenant service, and
may run out of resources  (CPU time) required for NDP processing.
Instead of waiting for resources and blocking progress, a Page Store
can skip NDP processing and just return the requested page.
In that case, {InnoDB} will complete the NDP processing of the page. 
As a result, the query executor can rely on the requested 
NDP processing being done---either by Page Stores or by {InnoDB}.
 
Because of multi-versioning, the latest version of a row on a page may
not be visible to a scan. A Page Store is unable to traverse a row's
undo chain and reconstruct older versions because the required
undo records may reside in other Page Stores. Such invisible rows
must be returned to {InnoDB}, which \textit{is} able to reconstruct
the correct older version, and perform the requested NDP
processing on the row.

\subsection{NDP support in the query optimizer}
\label{ndp_support_in_qo}
We considered two approaches to integrate NDP support into query optimization.

\begin{itemize}
    \item During plan enumeration, consider NDP as an
    alternative, and estimate its cost and benefits. This may influence
    join order, join types, table access methods, and so on.
    \item Treat NDP as a query plan post-processing step: finalize a query plan without considering NDP,
    and then consider enabling NDP for each of the table accesses in the plan.
\end{itemize}

The first approach  can potentially produce a faster plan,
but increases the optimizer’s search space.
For example, a hash-join with NDP pushdown may be better than a nested-loop join without NDP.
By not considering NDP during plan enumeration, the optimizer may miss the hash join plan. 
However, we opted for the second approach for several reasons.
\begin{enumerate}
    \item It does not require changing the core optimizer, and
just requires adding a (less intrusive and less risky)
post-processing step.
   \item Where possible, the optimizer already pushes down data
reduction operators on top of index scans. NDP is essentially
a more efficient way of evaluating the operators, so
the risk of performance regression is low.
   \item NDP processing is not guaranteed; Page Stores may ignore the 
optimizer's NDP request due to resource constraints.
\end{enumerate}

The post-processing approach works as follows.
For each table access in the final plan, 
the optimizer considers NDP column projection
and NDP predicate evaluation. For the last table access in a query block,
the optimizer also considers NDP aggregation if a
\texttt{GROUP BY} clause or
aggregation functions are present.
If the optimizer enables any of the three NDP features,
the table access is marked as an `NDP scan'.

NDP is only beneficial if an access method reads many rows:
for example, a full table scan or range index scan.
Accordingly, NDP is not considered for table access methods that
access only a few rows---for example, a point lookup.

\subsection{NDP support in InnoDB storage engine}
\label{ndp_support_in_innodb}
The {InnoDB} storage engine handles all of the complexities
related to NDP scans, and shields the SQL executor from NDP.
Indeed, the SQL executor only provides {InnoDB} the
necessary callback functions for predicate evaluation
or value accumulation (for aggregations).

\subsubsection{NDP Descriptor}
For an NDP scan, {InnoDB} encapsulates and builds all of the
relevant information in a data structure called an `NDP descriptor'.
A separate NDP descriptor exists for each table in a
query block, and contains the following information.
\begin{itemize}
    \item the number and data types of the index columns and the lengths
    of the fixed-length columns
    \item the columns to be projected, if any
    \item the encoded filtering predicates in the LLVM IR
    (intermediate representation) format, if any
    \item the aggregation functions to call
          and the \texttt{GROUP BY} columns, if any
    \item a transaction ID that represents an MVCC (multi-version
    concurrency control) read-view low watermark. If the transaction ID of a row is less than this low watermark, the row is visible to the scan; otherwise it may not be. Note that a complete list of active transactions is not included to reduce CPU overhead in Page Stores. 
\end{itemize}

\subsubsection{`NDP' Pages}
An NDP I/O request reaches a Page Store, and using
the accompanying NDP descriptor, the Page Store
converts a regular {InnoDB} page into an `NDP' page.
Unlike a regular fixed-length {InnoDB} page (usually 16 KB),
an NDP page is of variable length.
To avoid drastic code changes in {InnoDB}, and to use 
the same {InnoDB} code path to process both regular and
NDP pages, we decided that an NDP page should resemble a
regular {InnoDB} page.

\begin{itemize}
    \item An NDP page contains the same page header as a regular {InnoDB} page.
    The records in an NDP page have the same structure 
    as regular 
    InnoDB records. As a result, the existing InnoDB page cursor functions, which iterate over records in a page, remain unchanged. The code that formats a record (to extract fields from a record) can be used on the NDP record with 
    minimal changes.
    \item The records in an NDP page are also chained in index
    key order. 
    If a query uses the index to satisfy an ordering requirement, an
    NDP scan of the index still satisfies the ordering requirement,
    and a sort is avoided.
    \item As an optimization, if NDP predicate filtering removes all
    of the records in a page, the resulting empty page is indicated
    specially without requiring explicit materialization.
\end{itemize}

Although an NDP record resembles a regular {InnoDB} record,
there may be two differences. First, the NDP record may be narrower
because some columns may have been removed. Second, the NDP record
may represent an aggregation of multiple regular records.
A mix of regular records and NDP records can co-exist in an NDP page.

The {InnoDB} record header contains a ``record type'' field
which is reused to tag NDP records as indicated in
Listing~\ref{lst:ndp_record_types}. The two new status values
indicate to {InnoDB} row scan functions whether NDP projection or aggregation has happened on a particular record.\footnote{NDP
filtering removes records altogether, and therefore, does not
require a code.}
For regular InnoDB records, the scan functions follow the existing code path; for NDP records, NDP-specific code is used.

\begin{flushleft}
\begin{minipage}{\linewidth}
\begin{lstlisting}[basicstyle=\footnotesize,frame={tb},language=C,breaklines=true,caption={Two newly added NDP record types in storage/innobase/rem/rec.h},label={lst:ndp_record_types}]
#define REC_STATUS_ORDINARY 0
#define REC_STATUS_NODE_PTR 1
#define REC_STATUS_INFIMUM 2
#define REC_STATUS_SUPREMUM 3
#define REC_STATUS_NDP_PROJECTION 4
#define REC_STATUS_NDP_AGGREGATE 5
\end{lstlisting}
\end{minipage}
\end{flushleft}

\subsubsection{Interaction between NDP and the InnoDB buffer pool}
The existing {InnoDB} buffer pool is used to store NDP pages. Using the buffer pool to store both regular pages and NDP pages has the advantage of memory sharing. When there are no NDP scans, the entire buffer pool is still available to regular scans. Because NDP pages are custom made for a particular table access, although the NDP pages reside in the buffer pool, they should only be visible to the thread that performs the NDP scan and not to the other concurrent queries and transactions.\footnote{The converse is not true.
Regular non-NDP pages \textit{are} available to NDP threads: they
are simply copied to the NDP area of the buffer pool, and do not
require I/O's.} To achieve this invisibility, NDP pages are not inserted
into such buffer pool management data structures as hash map, LRU list,
flush list, etc. 
NDP pages are managed by {InnoDB} persistent cursors--—{InnoDB}’s regular mechanism for driving table access. The {InnoDB} persistent cursor is responsible for allocating NDP pages from the buffer pool free list, and releasing the NDP pages. The number of NDP pages allocated is controlled so that regular scans are not deprived of memory.

\subsubsection{NDP scans and batch reads}
Like a regular scan, an NDP scan also traverses a B+ tree
to locate leaf pages. When the traversal reaches a level-1 page
(the level immediately above the leaf level), the NDP scan extracts
the child leaf page ID's from the level-1 page, and packs
the leaf page ID's into a single I/O request, called a `batch read.'
A batch read’s memory footprint is known (controlled using a
newly introduced MySQL parameter called
$innodb\_ndp\_max\_pages\_look\_ahead$),
and an NDP scan’s memory footprint is set to be the same value:
after an NDP scan finishes processing an NDP page in the batch,
the page is immediately released back to buffer pool free list.

An NDP batch read uses page locking \textit{and} LSN versioning
for concurrency control as follows. Traditionally, page locking
is used to solve concurrent read-write conflicts in a
B-tree traversal. However, given an NDP batch read's
large size (typically around a thousand pages),
it impractical to lock and block modifications to
individual pages. During a B-tree traversal,
shared page locks are obtained starting from the root page
until a level-1 page. Since the sub-tree is share-locked,
no transaction can modify the sub-tree structure
(e.g., insert or delete a page).
Then an LSN (Log Sequence Number) corresponding to the
locked sub-tree structure is generated.

The LSN accompanies the NDP batch read request to the Page Store.
Once an NDP batch read request is submitted, the
the B-tree locks can be released, and the sub-tree may
be modified. The Page Store only returns those page
versions matching the LSN value, and thus, the batch read is
shielded from the concurrent B-tree modifications.

Before a leaf page ID is added to a batch read request, a
check is made whether the page already exists in the buffer pool.
If so, an I/O is avoided by copying the cached (non-NDP) page
to the NDP page area. A copy is required instead of using
the non-NDP page
directly because the page may be modified by concurrent transactions once we release the page locks, and we need to ensure the NDP scan observes a consistent sub-tree structure. Only those page ID's
not found in the buffer pool get inserted into a batch
read request.

A batch read is aware of scan boundaries. For
example, in an 
index range scan of $c1 \leq 1000$, where $c1$
is the index key, the batch read will not read leaf pages
beyond the range because level-1 pages store `boundary'
$c1$ values.

In addition to reducing the number of I/O requests,
batch reads offer two other benefits.
\begin{itemize}
    \item They facilitate parallelism in the Page Stores.
    A Page Store can assign a thread to work on a page,
    and multiple threads can process the batched pages in parallel.
    Large batch read sizes (around a thousand pages) also means
    that multiple Page Stores are likely to engage in servicing
    the request.
    \item They facilitate cross-page aggregation in Page Stores,
    details of which are provided in Section~\ref{ndp_aggregation_impl}.
\end{itemize}

NDP can be enabled in parallel index scans, as described in Section~\ref{pq_and_ndp}.

\subsection{NDP support in Page Stores}
\label{ndp_support_in_page_stores}
Because Page Stores are intended to support several frontend systems, including MySQL,
PostgreSQL, and openGauss, 
the NDP framework for Page Stores is DBMS-independent. 
DBMS-specific shared libraries can be loaded as plugins
into the Page Stores. The Page Store NDP framework accepts
an NDP descriptor as a type-less byte stream, 
which an NDP plugin interprets.
An NDP I/O begins as a regular page read returning 
a regular page that the NDP plugin then converts into an NDP page. 
Multiple threads undertake NDP processing of pages
concurrently, independently,
and in any order enabling 
flexibility and parallelism in the Page Store. 
The logical page ordering is enforced in the frontend storage
engine, not in the Page Stores.

Once a regular page has been read, an NDP plugin
iterates through the records in the page,
and checks whether a record is visible by comparing 
the record’s transaction ID  with the transaction ID in the NDP descriptor.
If the former is lower, the record is visible to the transaction
requesting the page; otherwise, the record is
\textit{ambiguous} in that
the Page Store cannot determine if it is visible.
Visible delete-marked records are skipped.
Such NDP operations as column projections, predicate evaluations,
and aggregations are then applied to the visible records.

A Page Store is a multi-tenant service that simultaneously
supports multiple frontend instances,
processing a mix of NDP and non-NDP read requests.
When there are many concurrent NDP requests,
a Page Store CPU may become a
bottleneck, and negatively affect the overall client response time.
To alleviate, two optimizations---NDP descriptor cache
and resource control---are introduced.

\subsubsection{NDP descriptor cache}
Initial performance tests revealed that NDP descriptor
decoding caused a bottleneck in Page Store CPU---a
few milliseconds per decoding on average---and slowed
down queries.
Significant CPU time was also spent compiling  LLVM bitcode
into native code and obtaining the function pointer. 
A typical query scanning a large table generates
many waves of NDP page read requests with the same NDP
descriptor to a Page Store.
This access pattern was leveraged
by introducing an NDP descriptor cache. 
Instead of decoding descriptors and converting LLVM bitcode for each NDP request, the first request caches the result
which is reused subsequently.
(The cache key is computed by applying a
hash function to the NDP descriptor fields.)
This optimization
dramatically reduced the average decoding time to less than 5
microseconds, and improved performance on
some benchmarks by up to 50\%.

\subsubsection{NDP resource control}
A Page Store keeps pages up to date by applying log
records, and serves page read requests.
It also performs several secondary tasks: compaction;
creating snapshots; and doing backups. 
Hence, it must be able to limit
the resources used for NDP requests. 
A dedicated thread pool was introduced to control the number of 
NDP pages processed concurrently. New NDP page read requests are
added to a queue, and wait for their turn. NDP processing
does not block regular page reads/writes,
and is treated as a best-effort activity. 
If the Page Store has enough resources to complete an
NDP request without undue 
waiting, the NDP processing of a page is done; otherwise, it is skipped, 
and the frontend node completes it. NDP resource control 
works closely with other Page Store flow control mechanisms to provide balance and fairness among different Page Store tenants.
Interestingly, because NDP resource
control is page-scoped, NDP benefit to a query is not
all-or-nothing: some pages might undergo NDP processing
before resource throttling kicks in, and NDP processing
is left to the {InnoDB} layer.

\section{NDP implementation}
\label{ndp_implementation_section}
NDP reduces data by retaining only the necessary
rows and columns required in a query, and by
aggregating the retained rows.
This section describes the implementation details of how
the NDP system performs column projection
(Section~\ref{ndp_col_proj_impl}); row filtering
using predicate evaluation
(Section~\ref{ndp_predicate_eval_impl}); and
row aggregation (Section~\ref{ndp_aggregation_impl}).

\subsection{NDP column projection}
\label{ndp_col_proj_impl}
For each table, the query optimizer estimates the total width of 
the columns required in a query, and compares it to
the total width of all of the columns. 
When the width reduction is high enough,
the query optimizer enables NDP column projection
for the table access.
For fixed-sized columns, the column widths can be easily obtained from the system dictionary. For variable-sized columns, average sizes---calculated using table statistics---are used.

In addition to the columns required by a query, some
fields needed by {InnoDB}’s internal processing are always included.
For example, the primary key columns are always included even if the query does not require them because {InnoDB} needs them for persistent cursor re-positioning. The transaction ID is also included for MVCC handling.

Only visible records are projected.
Ambiguous records are returned unchanged because {InnoDB} requires
the entire record to construct the old record version using its `undo' log.
Sending a `narrower' ambiguous record
could cause {InnoDB} to malfunction if the record is actually not visible, and
{InnoDB} needs to construct an older version.

\subsection{NDP predicate evaluation}
\label{ndp_predicate_eval_impl}
\subsubsection{NDP predicate evaluation workflow}
\label{ndp_predicate_eval_workflow}
Even without NDP, {MySQL}'s query optimizer always pushes 
down predicates into a table access when possible
(the `classical' predicate pushdown).
Only such pushed predicates are eligible for NDP evaluation;
cross-table predicates are not.

Not all data types and operators are supported by the LLVM engine
in Page Stores (Section~\ref{ndp_llvm_impl}), and
expressions with user-defined functions cannot be NDP-pushed
because they might pose security risks. The optimizer takes
a conservative approach, and maintains explicit lists
of allowed data types, operators, and functions. 
The optimizer then calculates the filter factors of the
predicates, and
enables NDP only if the predicates are sufficiently selective.
The query optimizer then separates NDP predicates from the original
ones: the residual non-NDP predicates are evaluated by the SQL executor.

Although the SQL executor never evaluates NDP predicates,
{InnoDB} may do so (by calling SQL executor functions) in
the following four cases.
\begin{enumerate}
    \item {InnoDB} handles ambiguous records---records that cannot be handled by
    Page Stores.
    \item A Page Store may not evaluate NDP predicates
    because of resource constraints, and {InnoDB} finishes the job.
    \item {InnoDB} may not even push NDP predicates to
    Page Stores because of its own resource constraints (buffer-pool pressure).
    \item An NDP page is copied from an existing (non-NDP) page in the buffer pool.
\end{enumerate}

A Page Store's NDP plugin invokes the LLVM engine to evaluate
NDP predicates on the records---as explained in
Section~\ref{ndp_llvm_impl}. A Page Store can only
safely discard `false' visible (unambiguous) records: for the
rest, decision must be deferred to {InnoDB}. 

Records disqualified by the NDP predicates are
removed from the page, and column projection is
performed on the surviving records, if applicable.

\subsubsection{The role of LLVM in predicate evaluation}
\label{ndp_llvm_impl}
Prior research has shown that interpretive expression
evaluation, as done by traditional relational systems,
can be slow~\cite{monetdb_paper, llvm_postgresql,
llvm_postgresql2, tpch_analyzed}. Therefore, Taurus
compiles expressions into bitcode---wrapped
in a function---and then calls the function once for each row.
In some of the prior research, LLVM query engines were built from
scratch~\cite{llvm_neumann, llvm_menon, llvm_klonatos, llvm_kohn}.
In contrast, Taurus LLVM bitcode compilation is non-invasive; requires
no changes to the existing Volcano-style SQL executor; combines
LLVM interpretation and execution; and uses a shared library of pre-compiled
complex functions. Bitcode for
predicates is generated just before query execution.

Classical (non-LLVM) {MySQL} predicate evaluation
proceeds by traversing a tree of various expression nodes,
and calling the necessary functions such as `$>$' and `$\leq$'.
This approach is slow because of the 
frequent function calls and cache misses.
LLVM, in contrast, traverses an expression tree bottom-up;
emits bitcode along the way; and creates a composite function
that encodes the entire expression tree.
This process is illustrated using the \texttt{WHERE} condition
``($a > 1$ \texttt{AND} $b > 2$) \texttt{OR} $c >= 3$''. The resulting IR (intermediate
representation) code appears in Listing~\ref{lst:llvm_ir_code}.
In the code, a label with the prefix `\%' represents an LLVM register.

\begin{flushleft}
\begin{minipage}[c]{\linewidth}
\texttt{AND} $b >= 2$) \texttt{OR} $c >= 3$.
\begin{lstlisting}[basicstyle=\footnotesize,frame=tb,language=LLVM,breaklines=true,caption={LLVM intermediate representation (IR) code for the predicate ``($a > 1$ \texttt{AND} $b > 2$) \texttt{OR} $c >= 3$''.},label={lst:llvm_ir_code}]
define i32 @f() #0 {
entry:
  %0 = load i32, i32* %a, align 4
  %cmp = icmp sgt i32 %0, 1 ; a > 1?
  br i1 %cmp, label %b_and_cont, label %b_or_cont ;shortcut may happen

b_and_cont:
  %1 = load i32, i32* %b, align 4
  %cmp1 = icmp sgt i32 %1, 2 ; b > 2?
  br i1 %cmp1, label %b_and_true, label %b_or_cont

b_and_true:
  store i32 1, i32* %retval
  br b_ret;

b_or_cont:
  %2 = load i32, i32* %c, align 4
  %cmp3 = icmp sge i32 %2, 3 ; c >= 3?
  store i32 %cmp3, i32* %retval
  br b_ret;

b_ret:
  ret i32 %retval
}
\end{lstlisting}
\end{minipage}
\end{flushleft}

LLVM execution requires several common utility functions---for example,
\textit{bin2decimal} that converts a decimal number's binary
representation into a format used by {MySQL}.
Such utility functions are pre-compiled,
and collected into a shared library that is installed on all of the
Pages Stores. This design choice eliminated the need to convert large complex functions into LLVM bitcode.

\begin{figure}
  \centering
  \includegraphics[height=3.4in,width=\linewidth]{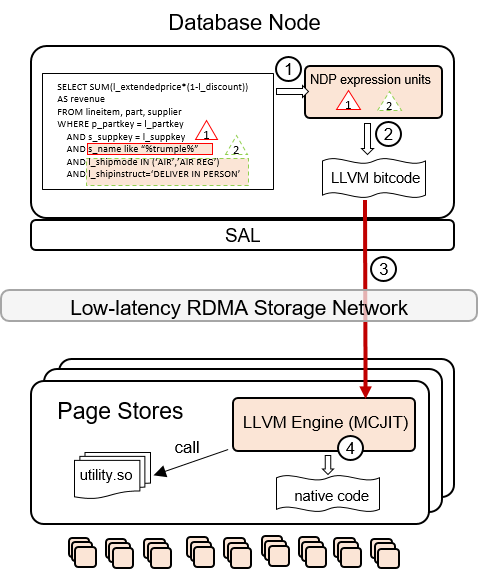}
  \caption{The four steps in the LLVM compilation workflow.}
  \label{fig:llvm_workflow}
\end{figure}

LLVM compilation itself consists of several steps depicted in
Fig.~\ref{fig:llvm_workflow} and described below.
\begin{enumerate}
    \item Predicates for each table in the query are identified
    and translated using the LLVM C/C++ API so that code generation can begin.
    As already indicated in
    Section~\ref{ndp_predicate_eval_workflow},
    predicate identification is done by the query
    optimizer based on the predicate's estimated selectivity.
    In Fig.~\ref{fig:llvm_workflow},
    table-scoped conditions marked as
    the triangles `1' and `2' are chosen,
    but cross-table conditions are left alone
    because they will not participate in NDP processing.
    \item Rewritten predicates are compiled into IR
    by the LLVM frontend Clang~\cite{clang}. To facilitate debugging
    and to identify mistakes, in-memory IR can be optionally persisted on disk.
    In Fig.~\ref{fig:llvm_workflow}, the
    expressions corresponding to the 
    triangles `1' and `2' are traversed bottom-up, and
    the IR code is emitted along the way as
    illustrated in Listing~\ref{lst:llvm_ir_code}.
    \item The resulting in-memory IR code
    is packed into the NDP descriptor and sent to
    each Page Store.
    There is a separate NDP descriptor per table.
    The IR codes
    for `1' and `2' are put in the NDP descriptors
    of the \textit{Supplier} and \textit{Lineitem} tables, respectively.
    \item A Page Store extracts the IR bitcode from the NDP descriptor
    and sends it to the LLVM execution engine. The engine returns the
    address of a function $f$ that encodes the predicates. The
    Page Store uses $f$---which may call some utility functions
    present in the shared library---to perform record filtering.
    To further speed-up $f$, it is just-in-time compiled into
    native machine code before the first call.
    Just-in-time compilation permits 
    architecture-specific native code generation---for example,
    ARM or X86---depending on the Page Store hardware.
    In Fig.~\ref{fig:llvm_workflow}, the IR codes for
    `1' and `2' are just-in-time compiled to native
    functions $f_1$ and $f_2$, which then filter \textit{Supplier}
    and \textit{Lineitem} rows, respectively.
\end{enumerate}

Care must be taken to ensure that filtering
and expression evaluation on Page Stores produce the
same result as that produced by the 
hypothetical non-NDP evaluation on the SQL node; 
otherwise, the query may produce incorrect results
because of arithmetic overflow, underflow, and floating-point arithmetic issues. 

\subsection{NDP aggregation computation}
\label{ndp_aggregation_impl}
{MySQL} query optimizer enables NDP aggregation on
a table $T$ based on the following logic.
\begin{itemize}
    \item {MySQL} query execution proceeds block-by-block,
    and therefore, $T$ must be the last table accessed in a
    query block. Furthermore, there must be no residual
    predicates that need still need evaluation by the SQL
    executor during or after the table access.
    \item If the aggregation is for a \texttt{GROUP BY} clause,
    then the index access chosen for $T$ must satisfy the
    grouping column requirement. This restriction exists
    because sort- or hash- \texttt{GROUP BY} is not implemented in Page Stores.
\end{itemize}



Page Stores perform aggregations on per-page basis, and
the work is best explained using an example.\footnote{In
this section, we assume that NDP predicate filtering---which
precedes NDP aggregation---has already happened inside
a Page Store.}

\begin{itemize}
    \item Suppose that an aggregation group on the page $P_1$
    has 5 records, and the aggregation
    function itself is \texttt{SUM}. Let
    $P_1 = \{(1, 2), (2, 10)?, (3, 7), (4, 8)?, (5, 2)\}$
    in which the first tuple value indicates record ID,
    and the second tuple value indicates the column value
    to be summed up.
    Two of the records are ambiguous in the sense
    described in Section~\ref{ndp_support_in_page_stores},
    and are denoted by `?'. Recall that the Page Stores
    cannot process ambiguous records.
    \item A Page Store computes $\mathrm{NDP}(P_1)$---the NDP-processed
    version of $P_1$---as follows. Visible (non-ambiguous)
    records---except the last record in a group---are summed
    up, and discarded; and the summation is attached to the last
    record. Thus, 
    $\mathrm{NDP}(P_1) = \{(2, 10)?, (4, 8)?, ((5, 2), 9)\}$ in which
    $9$ resulted from $2 + 7$, and the resulting longer
    record $((5, 2), 9)$ is an `NDP' aggregation record
    indicated with the value $5$ in
    Listing~\ref{lst:ndp_record_types}.
    \item In general, $P_1$ will have records with many
    grouping values, and each grouping value is handled similarly.
\end{itemize}

Page Stores can also aggregate across pages, and
there are two cases to consider.
\begin{enumerate}
    \item If \texttt{GROUP BY} clause is present,
    only logically adjacent pages can be aggregated.
    Page Stores generally do not know the logical order
    of pages, and therefore, cross-page aggregation does
    not happen.
    \item If \texttt{GROUP BY} clause is absent (scalar
    aggregation), even logically non-adjacent pages can
    be aggregated, and cross-page aggregation happens.
\end{enumerate}

For cross-page aggregation, the Page Stores have to
recognize which pages belong to the same table access from
the SQL node. It is difficult (and may not be feasible) to collect such
pages from different I/O requests. Therefore, a simpler approach was
chosen: cross-page aggregation happens only to the pages of the same I/O request.
{InnoDB}'s batch reads play an important role here because they
enable cross-page aggregations.

Continuing with the previous example, suppose that
\texttt{GROUP BY} clause is absent, and the scalar
\texttt{SUM} aggregation spans across another page $P_2$.

\begin{itemize}
    \item Let
    $P_2 = \{(11, 10), (12, 2)?, (13, 5), (14, 9)\}$.
    \item $\mathrm{NDP}(P_2) = \{(12, 2)?, ((14, 9), 15)\}$.
    \item Cross-page aggregation across $P_1$ and $P_2$,
    denoted by $\mathrm{NDP}(P_1, P_2)$ is computed as
    follows. Ambiguous records are left alone; non-ambiguous
    records are summed up; and the value is attached to
    the latter of the two pages in the batch I/O request
    ordering. Assuming $P_2$ is that latter page, 
    $\mathrm{NDP}(P_1, P_2) = \{(2, 10)?, (4, 8)?, (12, 2)?, ((14, 9), 26)\}$ in which $26$ results from
    $2$ ($P_1$) + 9 ($P_1$) + 15 ($P_2$).
\end{itemize}

{InnoDB} performs the residual aggregation work on the ambiguous
records, and shields the SQL executor from NDP
aggregations in the following sense. 
Consider the `NDP' aggregation record $((14, 9), 26)$.
Its prefix $(14, 9)$ is a regular (non-NDP) record,
and it sent to SQL executor. {InnoDB} then calls
the SQL executor's appropriate aggregation function
(`sum' in this case), and provides the special value $26$.

\section{Three levels of parallelism}
\label{pq_and_ndp}
Query processing in the {MySQL} 8.0 community version is single threaded,
but a different group at Huawei has added parallel query (PQ) capabilities.
The initial implementation is limited in scope:  a table or range
scan can be range-partitioned into many sub-scans
that are processed in parallel by a pool of worker threads.
A sub-scan can be converted into
an NDP scan as described in Section~\ref{ndp_support_in_innodb}.

\begin{figure}[h]
  \centering
  \includegraphics[height=3.4in,width=\linewidth]{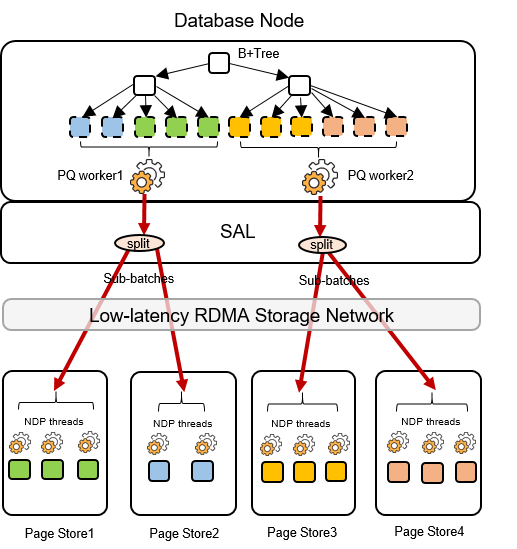}
  \caption{Three-level parallelism enabled by combining PQ and NDP.}
  \label{fig:three_level_parallism}
\end{figure}

By combining PQ and NDP, Taurus achieves three levels of parallelism 
as illustrated in Fig.~\ref{fig:three_level_parallism}: 
in the SQL node, across Page Stores, and within a Page Store.
These three levels of parallelism work 
together to reduce processing time significantly.

\subsubsection{SQL node parallelism}
PQ drives the SQL node parallelism; it partitions a table and uses 
multiple PQ worker threads to scan the partitions concurrently. 
The other two levels of parallelism are driven by
the NDP batch read capability. 

\subsubsection{Parallelism across Page Stores}
When a PQ worker
thread scans its assigned partition, the thread can activate
NDP, which sends batch reads to the Page Stores.
The pages in a batch are usually scattered across multiple slices,
and the slices are usually hosted by multiple Page Stores.
Specifically, the Storage Abstraction Layer (SAL) splits
a batch read into multiple sub-batches, based on where the
pages are located. Pages that belong to the same slice
are assigned to the same sub-batch. SAL concurrently sends
the sub-batches to Page Stores, with the effect that multiple
Page Stores are engaged in parallel to serve the original
batch read.

\subsubsection{Parallelism within a Page Store}
When a Page Store receives
a batch read request (which may be a sub-set of the
original batch read), the Page Store can use multiple concurrent
threads to serve the batch read, with each thread performing NDP
operations (column projections, predicate evaluations,
and aggregations) on its pages in the batch.

\section{Experimental results}
\label{experiments_section}

\subsection{Initial micro-benchmarks}
\label{network_read_reduction_section}
The most direct benefit of NDP is a reduction in network traffic:
data filtered out in Page Stores
never travels over the wire to {InnoDB} and beyond. This
effect can be clearly demonstrated using queries that simply count
the number of rows. The performance of \texttt{COUNT(*)} queries is a perennial
problem in {MySQL}, and NDP provides immediate customer benefits. 
We illustrate the gains on a 1 TB TPC-H database with a workload
consisting of the three \texttt{COUNT(*)} variants shown
in Listing~\ref{lst:count_star_variants},
plus Q1 and Q6 from TPC-H.

\begin{flushleft}
\begin{minipage}{\linewidth}
\begin{lstlisting}[basicstyle=\footnotesize,frame=tb,language=SQL,breaklines=true,caption={The \texttt{COUNT(*)} variants in the micro-benchmark.},label={lst:count_star_variants}]
Q0: SELECT COUNT(*) FROM lineitem;
Q001: SELECT COUNT(*) FROM lineitem
      WHERE l_shipdate < DATE '1998-07-01';  #table scan
Q002: SELECT COUNT(*) FROM lineitem
   WHERE l_suppkey <= 10000;  # secondary index scan
\end{lstlisting}
\end{minipage}
\end{flushleft}

The queries were run on a small test cluster with four Page Store nodes.
Each node was running on Intel\textsuperscript{\textregistered}
Xeon\textsuperscript{\textregistered} Gold 6161 CPU @ 2.20 GHz
with 44 cores, 250 GB memory, and had a Huawei Hi1822 network card
rated at 25 Gbps.
The SQL node had 360 GB of memory, but was otherwise identical
to the Page Store nodes.
Parallel query used 32 threads.

\begin{figure}[h]
  \centering
  \includegraphics[height=1.7in,width=\linewidth]{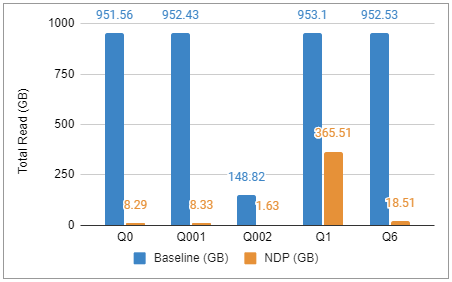}
  \caption{Network read reduction with NDP.}
  \label{fig:network_read_1tb}
\end{figure}

The plans for Q0 and Q001 use a table (primary index) scan,
and Q002 plan uses a secondary index scan. Q1 scans the
\textit{Lineitem} table and performs a \texttt{GROUP BY} with 
multiple aggregates.
Q6 computes one aggregate on the \textit{Lineitem} table,
but has several conjunctive predicates. As can be seen in
Fig.~\ref{fig:network_read_1tb}, with NDP, network reads are reduced to
negligible amounts for the \texttt{COUNT(*)} queries and Q6. The reduction is less for Q1 but is still considerable.

\begin{figure}[h]
  \centering
  \includegraphics[height=1.75in,width=\linewidth]{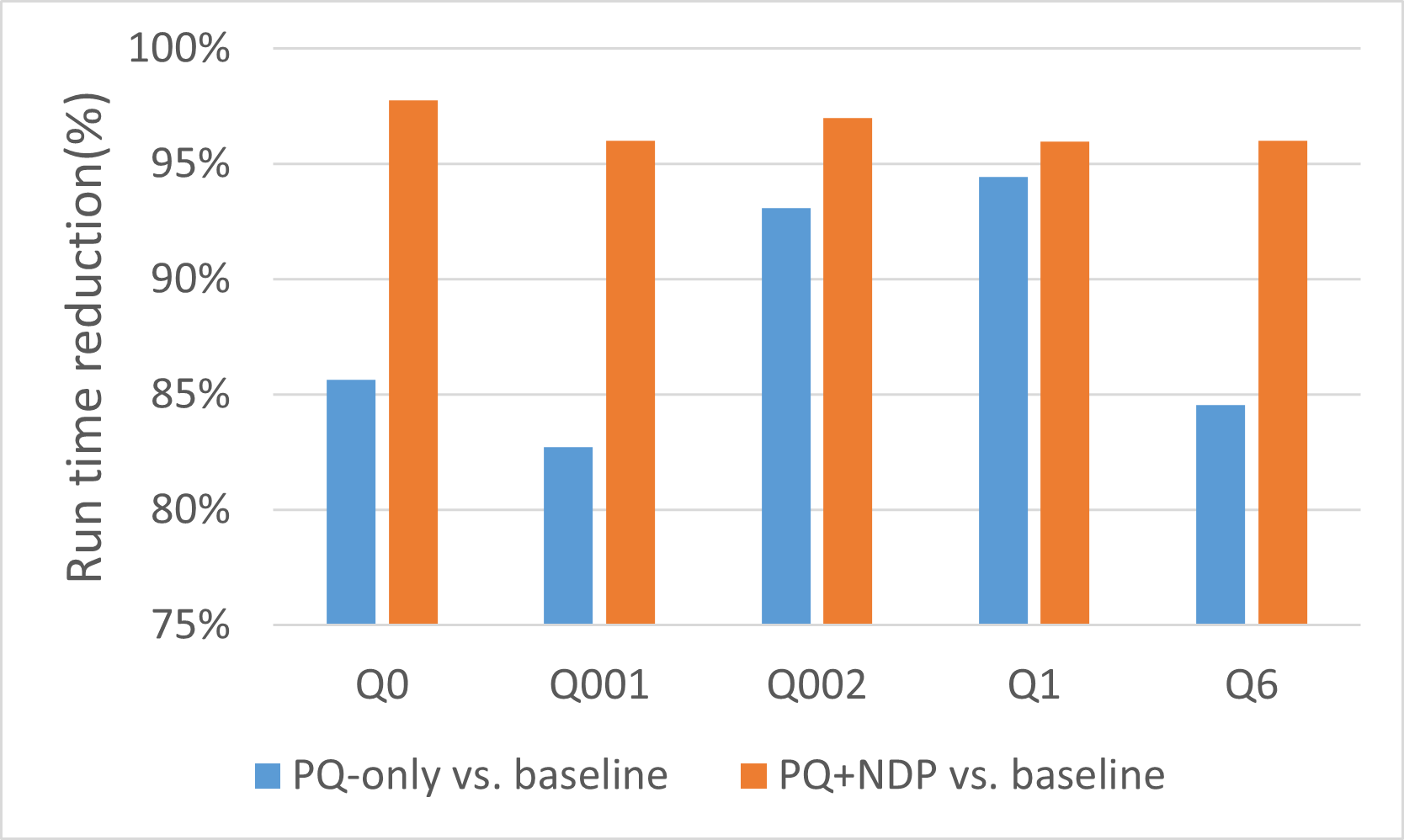}
  \caption{Run time reduction with NDP and PQ (higher is better).}
  \label{fig:runtime_1tb}
\end{figure}

Figure ~\ref{fig:runtime_1tb} shows the relative reduction in run time compared
with single-threaded execution without NDP or PQ, and
illustrates how NDP complements PQ.

With a PQ degree of 32, the theoretical run time reduction
is: $1-\frac{1}{32} = 96.875\%$.
However, with PQ only and no NDP, queries Q0, Q001, and Q6 achieve 
less than 86\% reductions because they must each transfer 
about 950 GB of  data over the network, and bottleneck on I/O.
Q002 and Q1 achieve relatively higher reductions with PQ-only
because they scan much smaller
secondary indexes, and are less I/O intensive.
Q1 is more CPU intensive than the other queries because of
its expensive aggregation expressions. 

When NDP is combined with PQ, we see further run time reduction
for all five queries. The reductions are all close to or achieve the theoretical maximum because with NDP enabled, the I/O bottleneck is avoided.

\subsection{TPC-H: Experimental Setup}
\label{tpch_100gb}
We ran the complete set of 22 TPC-H queries with and without NDP enabled
on a regular production cluster in Huawei's cloud
(Beijing region, instance class 16U64G).
The database size was 100 GB. 
The buffer pool size was set to 20 GB,
and the sort and join buffer sizes were both set to 1 GB.
We ran the 22 queries in sequence without restarting
the server in between.

\subsection{TPC-H: Data and CPU reduction}
\label{data_cpu_reduction}

\begin{figure*}[h]
  \centering
  \includegraphics[height=2in,width=\linewidth]{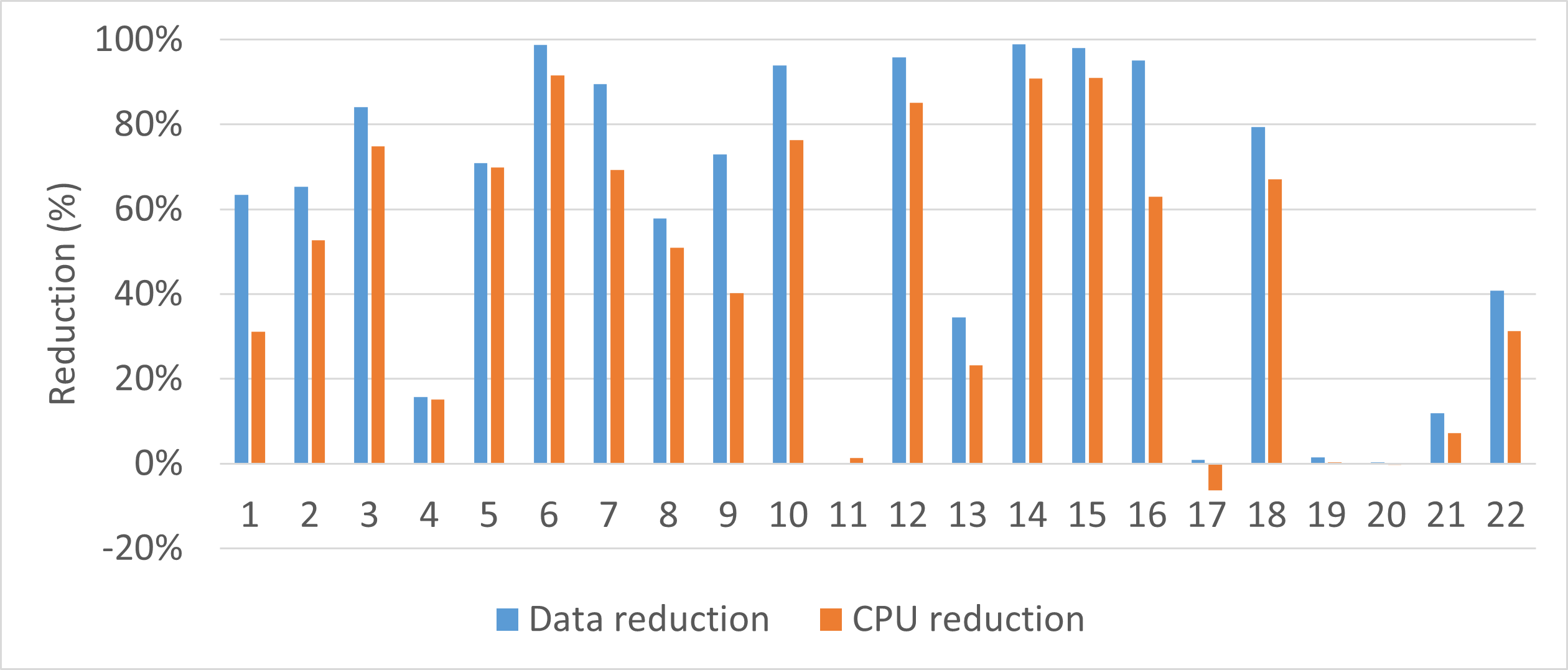}
  \caption{CPU time and network traffic reduction with NDP
           TPC-H queries.}
  \label{fig:savings_tpch}
\end{figure*}

Fig.~\ref{fig:savings_tpch} plots  the reduction
in network traffic (resulting in data reduction)
and CPU time on the SQL node with NDP enabled.
Overall, network traffic was reduced by 63\% and CPU time by 50\%,
and 18 out of the 22 queries benefited from NDP. 
In the following, we shall analyze a few queries in more
detail to gain insight into the factors influencing NDP's
effectiveness.

Queries Q6, Q12, Q14, and Q15 all exhibit over 90\% reduction in 
network traffic and over 85\% reduction in CPU time. 
They all have query plans that include scanning
the \textit{Lineitem} table where
filtering and column projection can be 
pushed down to Page Stores.
Q6 does nothing but scans the \textit{Lineitem} table and applies filtering
and aggregation. NDP achieves 99\% reduction of network traffic
and 91\% CPU reduction.
Q12 contains a hash join of \textit{Orders} and \textit{Lineitem} and applies
NDP to both inputs.
Q14 applies NDP on a scan of the \textit{Lineitem} table, and joins 
the remaining rows with \textit{Part} using an NL join,
achieving data and CPU reductions of 95\% and
89\%, respectively.
Q15 scans \textit{Lineitem}, applies NDP, and
achieves 98\% reduction in network traffic and 91\% CPU reduction.

Two queries, Q10 and Q16, also achieve over 90\% reduction
in network traffic but a slightly lower CPU reduction, 73\% and 63\%, respectively.

Queries Q11, Q17, Q19, and Q20 had plans with no NDP applied, and
consequently saw no reduction at all.
NDP is enabled on a scan only if the scan is estimated to cause
at least 10,000 pages of I/O.\footnote{
Just as an example, a scan size
based on table cardinality, row width, and selectivity might be
estimated at 14,000 pages, but at query run time, if 5,000
of the table's pages are in the buffer pool, only about 9,000 I/O's
can be expected, and the scan would \textit{not} quality for NDP.}
All four queries had plans where the only opportunities to apply NDP
were on scans that were deemed too small.
For Q11, the NDP-eligible scan was on the \textit{Nation} table.
For the other three queries, the NDP-eligible scans were on
the relatively small \textit{Part} table, and many of its pages remained 
in the buffer pool, so the scans were estimated to read
too few pages to qualify for NDP.
Out of those three, Q19 is chosen for further illustration.
Q19 performs a nested loop join on
\textit{Part} (outer
table) with \textit{Lineitem} (inner table) using the predicate
`$p\_partkey=l\_partkey$'. NDP did not happen on \textit{Part}
because of buffer pool caching; it did not happen on
\textit{Lineitem} because an index lookup on $l\_partkey$
provides an efficient access path, and on average, only 28 inner rows
are estimated to join with an outer row.

It is quite common for a query to require only a few columns from 
a table. For this reason, it may be beneficial to apply NDP
even when there is no filtering condition. 
Projection-only NDP was used in 8 of the 22 TPC-H queries yielding 
substantial benefits. On Q18, for example, projection-only NDP is
applied on two table scans (\textit{Orders}, \textit{Lineitem}) resulting in
a data reduction of 80\%, and CPU reduction of 67\%.
On Q9, it is applied on three scans (\textit{Orders}, \textit{Lineitem},
\textit{Partsupp}) achieving a data reduction of 62\%,
and CPU reduction of 42\%.

MySQL's current version of hash join  does not include
Bloom filter pushdown---a standard feature of most hash join
implementations---which would have allowed even further data reduction on the probe side of hash joins used in the query plans.

\subsection{TPC-H: Run-time reduction}
\label{run_time_reduction}
\begin{figure}[ht]
  \centering
  \includegraphics[height=1.75in,width=\linewidth]{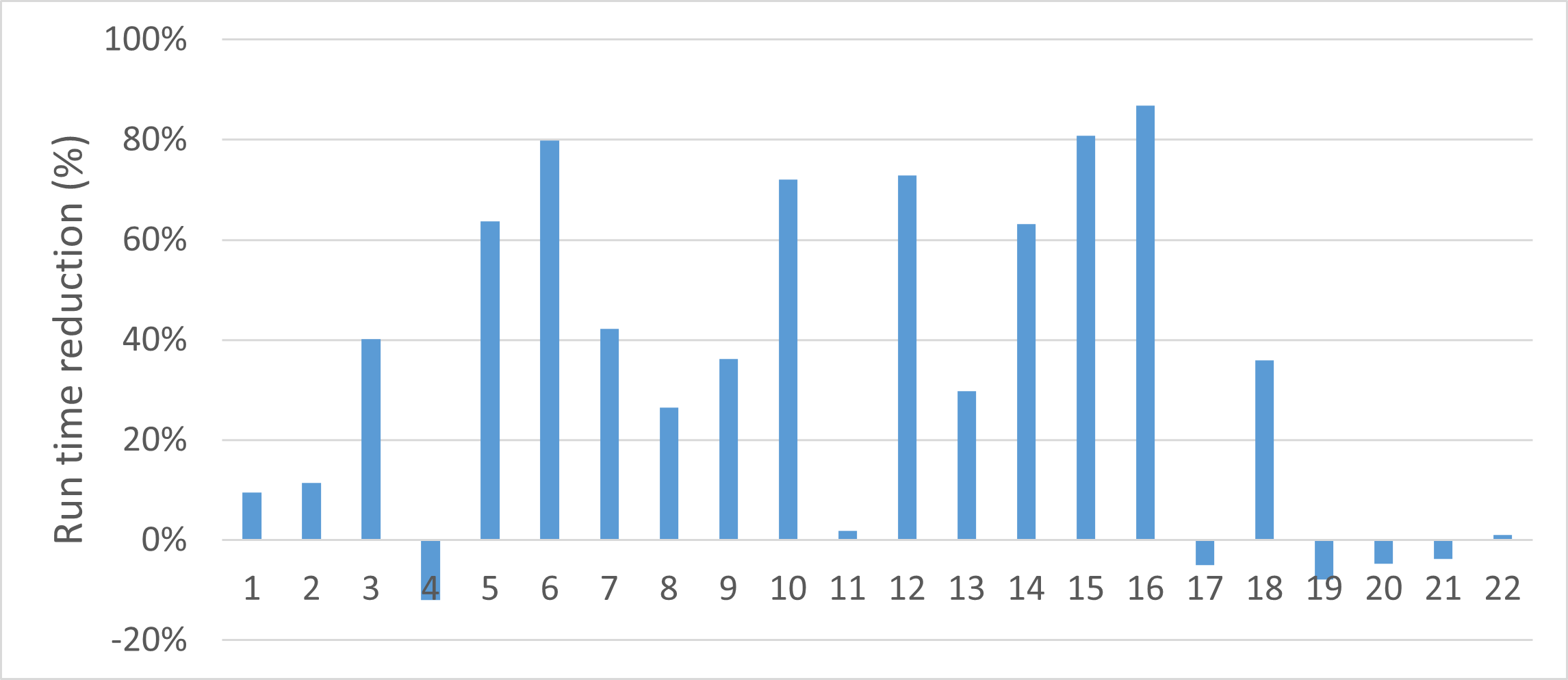}
  \caption{Run time reduction with NDP.}
  \label{fig:runtime_tpch}
\end{figure}

NDP delegates part of query execution to Page Stores, thereby
reducing the amount of processing performed on the SQL node.
This normally reduces query run time but not always, as we will see.
Fig.~\ref{fig:runtime_tpch} plots the relative reduction in
run time of the 22 TPC-H queries caused by NDP.
The total run time of the 22 queries was reduced by 28\%.
Run time was reduced by 60\% or more for
seven of the queries, and by as much as 80\% 
for three of the queries.

As expected, run time reduction is highly
correlated to data reduction: queries with the most
data reduction also tend to have the highest run time reduction.
However, Q4 is an apparent exceptions to this trend.
Q4 sees a data reduction of 16\% from NDP,
but run time increases by 12\%. 

Q4 performs a nested loop join of \textit{Orders} and
\textit{Lineitem} with \textit{Lineitem} as the inner. 
This generates a large number of lookups in the primary
index of \textit{Lineitem}, which is where most of the 
run time is spent. 
With NDP enabled, more of the lookups will cause 
a buffer pool miss because the three prior queries (Q1
through Q3) have not brought any \textit{Lineitem} pages
into the buffer pool.
Q2 does not access the \textit{Lineitem} table at all.
Q1 and Q3 do scan the \textit{Lineitem} table, but apply
NDP to the scans, so they do not bring any regular
\textit{Lineitem} pages into the buffer pool either.
So when Q4 runs with NDP enabled, it begins with
a buffer pool containing very few \textit{Lineitem} pages, 
resulting in a flurry of buffer pool misses.
If Q1 through Q3 ran with NDP disabled,
Q1 and Q3 would have brought \textit{Lineitem}
pages into the buffer pool, and Q4 would have
started with a `warm' buffer pool.
We verified this hypothesis with the following
experiment.
\begin{itemize}
    \item When Q1 through Q3 ran with NDP disabled, the
    resulting buffer pool had 1,272,972
    \textit{Lineitem} pages.
    \item When Q1 through Q3 ran with NDP enabled, the
    resulting buffer pool had only 24,186
    \textit{Lineitem} pages.
\end{itemize}

\subsection{TPC-H: Run-time further reduced by Parallel Query}
\label{pq_run_time_reduction}

\begin{figure}[ht]
  \centering
  \includegraphics[height=1.75in,width=\linewidth]{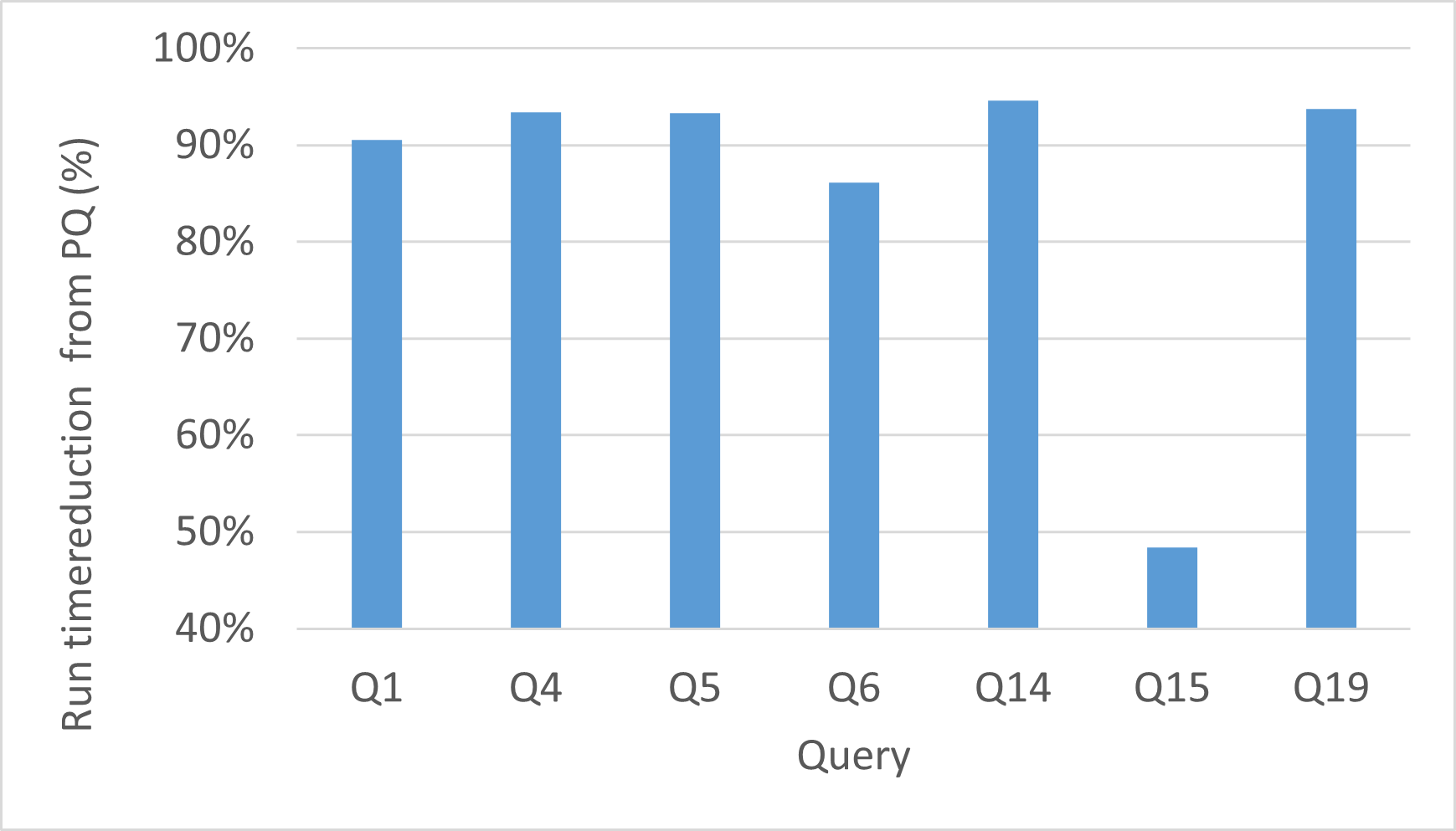}
  \caption{Further run time reduction from PQ.}
  \label{fig:runtime_ndp_pq}
\end{figure}

Parallel query can reduce run time of some but not all queries.
We repeated the test of TPC-H queries with both NDP and PQ enabled.
PQ reduced the run time further by at least 10\% 
on seven of the 22 queries.
Figure ~\ref{fig:runtime_ndp_pq} plots the additional run time reduction
(after NDP) from PQ on the seven queries.
The remaining queries saw no further reductions
because the optimizer chose fully serial plans. 
Huawei is currently enhancing PQ functionality to
enable parallelism in more queries.

The degree of parallelism was 16, so the maximum reduction is
$1-\frac{1}{16} = 93.75\%$.
On six queries, the run time reduction from PQ is close to the 
theoretical maximum.
However, for Q15, the reduction is only about half of the maximum.
NDP  achieves a data reduction
of 98\%, but the plan contains an NL join that is executed
serially, which limits parallelism gains.

Q1 contains an expensive aggregation operation
that is performed on the SQL node when run serially.
When PQ is enabled, this work is spread over
many worker threads, resulting in a substantial 
run time reduction over NDP alone.

Q4, Q5, and Q19 benefited from parallel NL joins:
multiple worker threads performing lookups on the inner
table(s) concurrently. 

\section{Related work}
\label{related_work_section}
Processing data near to where it lives is a decades-old
idea that  has seen increased
interest in the last $8$-$10$ years,
The research reported in
this paper is about near data processing
applied to database storage nodes, but as noted
in~\cite{ieee_micro_paper}, the same idea can be---and
has been---applied
at other levels of the memory hierarchy too: caches,
DRAM, nonvolatile storage-class memory, and so on.
For example, the `active-routing' suggested in~\cite{en_route_ndp}
pushes computation to a router attached to memory to 
better exploit parallelism and bandwidth of a memory bank.

A classification provided in~\cite{ndp_data_center_apps}
divides NDP into three categories based on 
the locations of the NDP-like operations.
\begin{enumerate}
    \item In-storage computing (ISC) for SSD-based approaches---sometimes
    also referred to as `SmartSSD' (SSD with an on-board
    FPGA)---for example~\cite{ndp_smart_ssds}.
    \item In-memory computing (IMC) for DRAM-based approaches---for
    example, the {JAFAR} accelerator described in~\cite{NDP_DaMoN}.
    \item Near-storage computing (NSC) for
    system-on-a-chip (SoC)-based approaches.
\end{enumerate}

The Page Store-based NDP processing of this paper is an ISC approach.
Out of those three categories, the FPGA-based ISC
approaches seem to have received the
most attention as detailed later in this section.
Indeed, some researchers are advocating
that time has come to create NDP-aware data centre servers
based on application needs: compute-intensive, data-intensive,
and possibly re-configurable varieties of them~\cite{ndp_data_center_apps}.

There are two fundamental reasons for the recent
surge of interest in NDP. First, big-data applications
need to process large data volumes, and information
extracted from such applications are often complicated
summaries, thereby offering aggregation and filtering
opportunities. More important, query optimizers can
push down aggregation and filtering---in many cases---to
such data containers as tables and indexes
residing on disk servers.
Second, in the increasingly common
cloud-deployed applications, disk servers are remote
even to their compute servers,
and early filtering saves network bandwidth
between the two before subsequently saving CPU cycles
on the compute servers.

In this research, the NDP decisions are taken by the
{MySQL} query optimizer, but as suggested in~\cite{ndp_ReProVide},
a disk server-resident local optimizer can optimize
selected operators,
gather data statistics, and cooperate with the global
optimizer. A prototype of such a system was
demonstrated using the Apache Calcite DBMS framework~\cite{apache_calcite}.

For analytical workloads, the benefits of equipping storage nodes with
computational power have long been understood, for example,
in Oracle Exadata~\cite{oracle_exadata_whitepaper} and {MySQL}'s
NDB cluster~\cite{mysql_ndb_cluster}.
In cloud-native database systems, separating
compute nodes from storage nodes has become standard. 

In addition to Taurus~\cite{taurus_paper}, two other MySQL-based
offerings separate compute and storage: Amazon's Aurora~\cite{mysql_amazon_rds},
and Alibaba's {POLARDB}~\cite{polardb_paper}.
Although all three systems separate compute
and storage, their design and implementation are very different.

In Taurus, NDP exploits parallelism in storage nodes, 
whereas PQ exploits parallelism in compute nodes. NDP and PQ work seamlessly together.

Aurora's `parallel query' feature is in fact a limited form of NDP: 
it pushes predicate evaluation and projection (but not aggregation) 
down into storage nodes~\cite{amazon_aurora_pq, amazon_aurora_pq_docs,amazon_aurora_pq_docs_group_by}.
Parallelism arises from the fact that there are multiple storage nodes
but processing in the compute node apparently remains
single threaded.
`Parallel query'  requires pushdown of at least
one \texttt{WHERE} clause (except for join queries),
whereas Taurus does not have that restriction.

In Taurus NDP, the compute node receives a single data stream
from storage nodes, and NDP processing is fully encapsulated within InnoDB.
In Amazon Aurora, the compute node receives two data streams from storage nodes: a “partial result stream” and a “raw stream.” The raw stream passes through InnoDB, the SQL execution engine, and finally lands in a PQ-specific component named “Aggregator”. The partial result stream bypasses InnoDB and goes to the Aggregator directly. The Aggregator combines the two streams into a single stream for further processing.

Alibaba's {POLARDB} ~\cite{polardb_pq} 
implements parallel query execution in the SQL node similar to Taurus,
but does not push data reduction operations to storage nodes, and
hence has no NDP support.
Reference ~\cite{polardb_paper} describes a joint pilot project aimed at
pushing table-scan operators into SSD drives
and implement the scans using {FPGA}. The project does not appear
to have gone beyond the pilot stage.

The recently announced {AQUA} project from
Amazon Web Services~\cite{amazon_aqua} takes a somewhat similar approach
by installing {FPGA} modules
next to the SSD's storing data, and then having
the FPGA's do data filtering, aggregation, compression,
and encryption.
A similar FPGA-based NDP approach for the RocksDB
key-value store is demonstrated in~\cite{ndp_kv_stores}, and
reports NDP benefits on point queries, range scans, and graph
analysis queries.
The `intelligent storage engine' described in
the Ibex prototype~\cite{ibex_paper} is also an NDP
engine, and can push down projections, selections, and grouping
operations. Because the implementation is FPGA-based,
each row of data read from a SATA disk is annotated with
its column metadata.

{Exadata}~\cite{oracle_exadata_whitepaper} `Smart Scans' perform row filtering
and column projection, but not aggregation in storage
nodes. It can handle filtering operations on compressed data.
Bloom filters computed during the build phase of hash joins
can also be pushed down. 
Exadata storage also maintains index-like structures (storage indexes) that help reduce physical I/O.

In Amazon Web Services' `S3 Select', selections,
projections, and scalar aggregates can be pushed
into S3 storage nodes~\cite{s3_select}. Data must be in
CSV, JSON, or Parquet~\cite{apache_parquet} formats.
Two recent prototypes PushdownDB~\cite{pushdowndb_paper}
and FlexPushdownDB~\cite{flexpushdowndb_paper} were
developed using `S3 Select'. The former pushed selections,
projections, and aggregates; the latter combined that
with data caching.

\section{Conclusion and Future Work}
\label{conclusion_section}
In Taurus,  
near-data processing (NDP) pushes data reduction operators 
(selection, projection, and aggregation) 
from the compute node to storage nodes (Page Stores), and reduces
data sizes close to the source. For analytical
queries, much less data travels over the wire from Page
Stores to the compute node. Less CPU processing on the
compute nodes may translate to reduced query run time.
Parallel query (PQ) deploys multiple threads to process
partitioned data which can further reduce run time.
As the experiments on TPC-H queries showed, the effects can
be dramatic: on Q15 
data shipped was reduced by 98\%, CPU time
by 91\%, and run time by 80\%.

NDP reduces CPU load on the compute nodes, and
the freed up CPU cycles become available to other queries,
enabling higher system throughput.
On the TPC-H queries, total CPU time on the compute nodes
was reduced by as much as 50\%.

The NDP implementation in Taurus affected three system
layers: query optimizer, {InnoDB} storage engine, and page stores.
We made conservative design choices that favored simplicity
over complexity; avoided cascading code changes; and minimized
chances of performance regressions.

Several directions for future work are possible.
NDP operations and parallel query execution need to be integrated
into the cost-based query optimization.
NDP expression evaluation needs to be extended
to support more data types and more operators.
The current NDP implementation only pushes down local predicates 
(predicates involving columns from a single table). 
We plan to push down join predicates in the form of Bloom filters.
Another possibility is to rewrite predicates
to make more of them NDP-eligible as was done in
Amazon Redshift~\cite{redshift_paper}
and AQUA~\cite{amazon_aqua}.

A separate team at Huawei is
planning to add NDP functionality to GaussDB for {OpenGauss}
by writing NDP plugins specific to that PostgreSQL-based system.

\bibliographystyle{IEEEtran}
\balance
\bibliography{IEEEabrv,sample-base}

\end{document}